\documentclass[pra,twocolumn,preprintnumbers,amsmath,amssymb]{revtex4}

\usepackage{graphicx}% Include figure files
\usepackage{dcolumn}% Align table columns on decimal point
\usepackage{bm}% bold math
\usepackage{color}
\usepackage{epstopdf}

\usepackage{ulem}
\def\bk{{\bm \kappa}}

\newcommand{\br}{ {\bm r}}

\def\bk{{\bm k}}

\def\br{{\bm r}}

\begin{document}

%\widetext

%\preprint{APS/123-QED}

%\title{Thermalization of random waves in the presence of strong disorder:\\
%Wave turbulence theory, simulations, and experiments in multimode optical fibers(hopefully...)}% Force line breaks with \\
%\title{Nonequilibrium condensation against strong disorder}% Force line breaks with \\
\title{Rayleigh-Jeans thermalization vs beam cleaning in multimode optical fibers}
%\title{Thermalization against equilibration mediated by strong disorder:
%Wave turbulence theory, simulations, and experiments in multimode optical fibers}% Force 
%\author{authors}
\author{K. Baudin$^{1}$, J. Garnier$^{2}$, A. Fusaro$^{3}$, C. Michel$^{4}$, K. Krupa$^{5}$, G. Millot$^{1,6}$, A. Picozzi$^{1}$}
%\author{Josselin Garnier$^{1}$, Kilian Baudin$^{2}$, Adrien Fusaro$^{2}$, Antonio Picozzi$^{2}$}
%\affiliation{$^{1}$ labs}
\affiliation{$^{1}$ Laboratoire Interdisciplinaire Carnot de Bourgogne, CNRS, Universit\'e de Bourgogne, Dijon, France}
\affiliation{$^{2}$ CMAP, CNRS, Ecole Polytechnique, Institut Polytechnique de Paris, 91128 Palaiseau Cedex, France}
\affiliation{$^{3}$ CEA, DAM, DIF, F-91297 Arpajon Cedex, France} 
\affiliation{$^{4}$ Universit\'e C\^ote d'Azur, CNRS, Institut de Physique de Nice, Nice, France}
\affiliation{$^{5}$ Institute of Physical Chemistry Polish Academy of Sciences, Warsaw, Poland}
\affiliation{$^{6}$ Institut Universitaire de France (IUF), 1 rue Descartes, 75005 Paris, France}


\begin{abstract}
Classical nonlinear waves exhibit, as a general rule, an irreversible process of thermalization toward the Rayleigh-Jeans equilibrium distribution. On the other hand, several recent experiments revealed a remarkable effect of spatial organization of an optical beam that propagates through a graded-index multimode optical fiber (MMF), a phenomenon termed beam self-cleaning. Our aim here is to evidence the qualitative impact of disorder (weak random mode coupling) on the process of Rayleigh-Jeans thermalization by considering two different experimental configurations. In a first experiment, we launch speckle beams in a relatively long MMF. Our results report a clear and definite experimental demonstration of Rayleigh-Jeans thermalization through light propagation in MMFs, over a broad range of kinetic energy (i.e., degree of spatial coherence) of the injected speckle beam. In particular, the property of energy equipartition among the modes is clearly observed in the condensed regime. The experimental results also evidence the double turbulence cascade process: while the power flows toward the fundamental mode (inverse cascade), the energy flows toward the higher-order modes (direct cascade). In a 2nd experiment, a coherent laser beam is launched into a relatively short MMF length. It reveals an effect of beam cleaning driven by an incipient process of Rayleigh-Jeans thermalization. As discussed through numerical simulations, the fast process of Rayleigh-Jeans thermalization observed in the 1st experiment can be attributed due to a random phase dynamics among the modes, which is favoured by the injection of a speckle beam and the increased impact of disorder in the long fiber system.
\end{abstract}

%\pacs{42.65.Sf, 05.45.a}

\maketitle

%Nonlinear optics in multimode fibers is an emerging field of research that has been attracting significant interest by the scientific community over the last few years. Among other phenomena, beam self-cleaning was discovered, which may enable new fiber optics technologies. Kerr nonlinearity phase-locks all fiber modes, thus generating at the fiber output high-quality and bright bell-shaped beams, instead of the usual speckled intensity patterns. Here, we present an original unified description of self-cleaning within a thermodynamic framework. Kerr beam cleaning is described as an equilibrium established among the multitude of modes, and it results from nonlinearity-induced thermalization, that is simply defined via conservation laws of statistical mechanics. Theoretical predictions are compared with extensive experimental studies, based on an new holographic technique, which permits us to compute the full mode content of beams emerging from the multimode fiber. Aiming at reaching out to the broad readership of Optics Express, we made particular efforts in describing self-cleaning within its broad picture, as provided by the current state-of-the-art. We believe that our results are of wide interest and will have significant impact, due to the current enthusiasm and rapid developments in the field. 

\section{Introduction}

Understanding the mechanisms responsible for self-organization processes in conservative wave systems is a difficult problem that generated significant interest. Contrary to a dissipative system, a conservative Hamiltonian system cannot evolve towards a fully ordered state, because such an evolution would imply a loss of statistical information that would violate the formal reversibility of the evolution equations. 
However, in spite of its formal reversibility, a nonintegrable Hamiltonian wave system is expected to exhibit an irreversible evolution toward the equilibrium state, i.e., the most disordered state that realizes the maximum of entropy \cite{sagdeev88}.
The {\it thermalization} to the Rayleigh-Jeans (RJ) equilibrium distribution of weakly nonlinear random waves can be described in detail by the well-developed wave turbulence theory \cite{zakharov92,nazarenko11}, that has been also successfully applied to optical waves \cite{turitsyn13,churkin15,PR14}.

RJ thermalization can be characterized by a counter-intuitive process of self-organization.
To discuss this aspect with some concrete examples, let us consider the nonlinear 
Schr\"odinger equation (NLS).
In the focusing regime, the self-organization process is known as ``soliton turbulence".
It is characterized by the spontaneous formation of a large scale coherent solitary-wave that remains immersed in a sea of small-scale fluctuations \cite{zakharov88,jordan,rumpf01,rumpf03,ZakhPhysRep01,
nazarenkoPR,nazarenko11,rumpfPRL09,PR14,PRA15,shalva}. 
%The solitary wave then plays the role of a ``statistical attractor" for the Hamiltonian system, while the small-scale fluctuations contain, in principle, all the information required for time reversal.
On the other hand, in the defocusing regime of the NLS equation, the self-organization process refers to `wave condensation'
\cite{Newell01,nazarenko11,Newell_Rumpf,PR14,PRL05,Nazarenko06,PD09,brachet11,OC12,
nazarenko14,turitsyn12,PRL18,NP12,bloch21}.
It originates in the singular behaviour of the RJ equilibrium distribution, in analogy with the quantum Bose-Einstein condensation.
More precisely, the RJ distribution exhibits a phase transition to a condensed state that is characterized by a macroscopic population of the fundamental mode of the system.

The observation of thermalization with optical waves in a (cavityless) free propagation is known to require very large propagation lengths, a feature that has been discussed recently in different circumstances \cite{chiocchetta16,PRL18}. 
The situation is completely different when the optical beam propagates in a waveguide configuration.
Indeed, the finite number of modes supported by the waveguide significantly increases the rate of thermalization to the RJ distribution \cite{PRA11b}.
In addition, the finite number of modes introduces a frequency cut-off that regularizes the ultraviolet catastrophe inherent to classical waves \cite{PRA11b,note}.
It is in this framework that a remarkable phenomenon of spatial beam self-organization, called `beam self-cleaning', has been recently discovered in graded-index multimode fibers (MMFs) \cite{krupa16,liu16,wright16,krupa17}.
%At variance with an apparently similar effect driven by the dissipative Raman effect in MMFs, known as Raman beam cleanup \cite{terry07}, this self-organization is due to a purely conservative Kerr nonlinearity \cite{krupa17}.
Recent works indicate that this phenomenon of beam cleaning can be interpreted as a consequence of RJ thermalization and condensation, a feature that has been discussed both theoretically \cite{PRL19,pod19,PRA19,christodoulides19,kottos20} and experimentally \cite{PRL20,fabert20,EPL21,wise_NP22,pod22,mangini22,PRL22,wright22}.
We recall in this respect that light propagation in an optical fiber is affected by a structural disorder due to refractive index fluctuations introduced by inherent imperfections and environmental perturbations \cite{mecozzi12a,mecozzi12b,mumtaz13,xiao14,psaltis16,faccio19}, 
which significantly impact the process of RJ thermalization \cite{PRL19,wabnitz_oft19,PRL22}.
We also note that RJ thermalization to negative temperature equilibrium states in multimode systems has been theoretically predicted \cite{christodoulides19,EPL21}, and recently observed in graded-index multimode optical fibers \cite{PRL23}.

In this work we report an accurate 
%and unequivocal 
experimental demonstration of the process of RJ thermalization through light propagation in a MMF. 
%and we show that it requires 
%We show that RJ thermalization requires random phases 
In a 1st experiment, we consider the injection of {\it speckle beams} in a relatively long MMF \cite{PRL20}.
We compare the experimental distribution of the power among the fiber modes with the corresponding theoretical RJ distribution, which reveals a  quantitative agreement without using adjustable parameters.
Furthermore, such a remarkable agreement is obtained over a broad range of kinetic energies $E$ (`temperatures'), i.e., over a broad  range of randomness of the speckle beams launched in the MMF.
In this way, we provide the experimental observation of the equipartition of the kinetic energy among the fiber modes.
The experimental results also evidence that, while the power (or `wave-action', or particle number in a corpuscular picture) flows toward the fundamental mode (inverse turbulence cascade), the energy flows toward the high-order modes (direct turbulence cascade).
Next, in a 2nd experiment, we consider the injection of a {\it coherent laser beam} in a short piece of MMF length, while the power is increased so as to keep constant the effective nonlinear interaction with respect to the 1st experiment.
%This configuration is similar to previous experiments reported in Refs.\cite{wise_NP22,mangini22}.
At variance with the 1st experiment, in the 2nd experimental configuration there are no random phases among the initially populated modes, while the impact of disorder on light propagation (polarization mixing and random mode coupling) is severely limited by the short fiber length used.
In this 2nd experiment, while we can identify an effect of beam self-cleaning driven by the thermalization of the low-order modes, the high-order modes exhibit a deviation from the RJ distribution.
%a feature that corroborate previous experimental results obtained in similar conditions.
%It reveals an effect of beam cleaning driven by the thermalization of the low-order modes, while the high-order modes exhibit a deviation from the Rayleigh-Jeans distribution.
%It reveals an effect of beam cleaning driven by an incipient process of Rayleigh-Jeans thermalization.
As will be discussed through numerical simulations, the fast process of Rayleigh-Jeans thermalization observed in the 1st experiment can be attributed due to a random phase dynamics among the modes, which is favoured by the injection of a speckle beam and the increased impact of disorder in the long fiber system.
%Finally, we discuss and interpret distinguished features that characterize the results of the two experiments.
%This work contributes to the understanding of the fast process of optical thermalization observed in multimode optical fibers and its interplay with the effect of spatial beam self-cleaning. 

\section{First experiment: Rayleigh-Jeans thermalization}

\subsection{Experimental setup and procedure}

%The experimental setup has been described in detail in Ref.\cite{PRL20}.
%Here we summarize the main characteristics.
Figure~1 schematically reports the experimental setup. The source is a Nd:YAG laser delivering subnanosecond pulses (400ps) at $\lambda_0=$1064 nm. 
%To prevent unwanted feedback into the laser cavity, we use an optical isolator and 
%We control the power with a half-wave plate and a polarizer. 
A key feature of the 1st experiment is that the laser beam is passed through a diffuser before the injection of the speckle beam into the MMF.
More specifically, the laser beam is collimated and passed through a glass diffuser plate placed in the vicinity of the Fourier plane of a 4f-optical system. 
%The beam is then launched into the MMF. 
%by using a lens $f_1$ = 15mm. 
%The near-field (NF) and far-field (FF) intensity distributions are measured at the fiber output following the procedure of Ref.\cite{PRL20}.  
The near-field (NF) intensity distribution of the output beam was magnified and imaged on a first CCD camera owing to a two lens telescope optical system.
% with $f_2$ = 8 mm and $f_3$ = 150 mm. 
The CCD camera was placed on a rail orthogonal to the beam propagation in order to remove or put the camera back on the beam path. 
The far-field (FF) intensity distribution of such a magnified image was obtained by placing it in the object focal-plan of a lens 
%$f_4$ = 250 mm 
and using a second CCD camera positioned in its image (Fourier) focal-plan, see Fig.~1.
%We refer the reader to Ref.\cite{PRL20}  for more details about the experimental setup.

An other important property of the 1st experiment considered in this section is that the MMF used is relatively long, i.e., $L=12$m long graded-index MMF.
Its  refractive index profile exhibits a parabolic shape in the fiber core with a maximum core index (at the center) of $n_{\rm co}$=1.470 and $n_{\rm cl}=1.457$ for the cladding at the pump wavelength of 1064nm (numerical aperture NA=0.195, fiber radius $R=26 \mu$m).
The corresponding parabolic potential reads $V(\br)=q |\br|^2$ for $|\br| \le R$, where $R=26 \mu$m is the fiber radius and $q=k_0(n_{\rm co}^2-n_{\rm cl}^2)/(2 n_{\rm co} R^2)$, $k_0=2\pi/\lambda_0$ the laser wave-number.
The MMF guides $M \simeq 120$ modes (i.e., $g_{max}=15$ groups of non-degenerate modes).
The truncation of the potential introduces a frequency cut-off in the FF spectrum 
$k_c=(2\pi/\lambda_0)\sqrt{n_{\rm co}^2-n_{\rm cl}^2}$.
The eigenvalues are well approximated by the ones of the ideal harmonic potential $\beta_p=\beta_0(p_x+p_y+1)$, where the index $\{p\}$ labels the two numbers $(p_x,p_y)$ needed to specify a mode. 
We have computed analytically the propagation of the optical wave throughout the setup of our detection scheme. 
The optical field amplitude in the NF plane is an exact magnification of the wave amplitude at the output of the MMF. In addition, the optical field amplitude in the FF detection plane exactly corresponds to the Fourier transform of the amplitude at the fiber output.
We note that the experimental setup for the NF and FF detection does not introduce detrimental spurious transverse phase profiles in the optical field, e.g., related to optical free propagation in air or phase shifts due to the presence of additional lenses.
For more details about the experimental set-up, we refer the reader to Ref.\cite{PRL23} (Supplementary Material). 

\begin{figure}
\includegraphics[width=1\columnwidth]{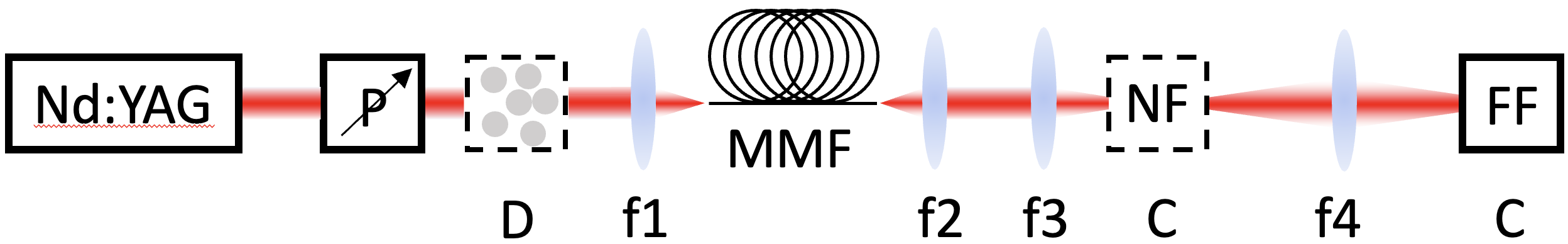}
\caption{
\baselineskip 10pt
{\bf Experimental setup:} laser, optical isolator, half-wave plate and polarizer, diffuser (D), lenses for magnification and imaging ($f_j$), graded-index MMF, and cameras for near- and far-field detections (C).
}
\label{fig:setup} 
\end{figure}

This setup allowed us to retrieve an accurate measurement of the power $N$ and the energy $E$ of the speckle beam.
%After propagation through a fiber length $L=12$m, both the near-field (NF) and the far-field (FF) intensity patterns are recorded with a (CCD) camera.
More precisely, the NF intensity distribution $I_{\rm NF}(\br)=|\psi|^2(\br)$ provides a measurement of the power $N=\int I_{\rm NF}(\br) d\br$ and of the potential energy $E_{\rm pot}=\int V(\br) |\psi|^2(\br) d\br$.
The kinetic energy $E_{\rm kin}=\alpha \int |\nabla \psi|^2(\br) d\br$ is retrieved from the FF intensity distribution $I_{\rm FF}(\bk)=|{\tilde \psi}|^2(\bk)$, where $\alpha=1/(2 n_{\rm co} k_0)$ and $\tilde \psi(\bk)$ is the Fourier transform of $\psi(\br)$.
%$k_0=2\pi/\lambda_0$ being the laser wave-number and $n_{\rm co}$ the refractive index.
This provides the measurement of the (linear) energy (Hamiltonian) 
$E=E_{\rm pot}+E_{\rm kin}$.
Projecting onto the basis of the fiber modes, the power and (kinetic) energy read
\begin{eqnarray}
N=\sum_p n_p, \quad E=\sum_p \beta_p n_p,
\end{eqnarray}
where $n_p$ 
%$n_p=|\int \psi(\br) u_p(\br) d\br|^2$ 
is the amount of power in the mode $\{p\}$, and $\sum_p$ is carried over the set of $M$ modes indexed by $\{p\}$ \cite{PRA11b}.
Note that $E$ is in units of W.m$^{-1}$ (not Joule). 
However, we refer $E$ to the energy because it is the (dominant) linear contribution to the Hamiltonian.
%According to the nonlinear Schr\"odinger (NLS) model, $E$ and $N$ are conserved quantities during the propagation in the MMF \cite{PR14,PRA19,horak}.

The accurate measurements of the NF and FF intensity distributions allowed us to retrieve the modal content of the optical beam in the experiment, namely the distribution of power among the modes $n_p^{\rm exp}$.
%We note in this respect that a full description of the beam includes a characterization of the amplitudes and phases of the waveguide eigenmodes, a problem known as mode decomposition (MD).
In general, to achieve the mode decomposition, several interferometric approaches based on use of a reference beam have been exploited to study light thermalization in MMFs \cite{wise_NP22,mangini22}.
%such as digital holography19,20 and multi-plane light conversion21,22 have
%been proposed, which require a coherent radiation source on the receiver.
%19. Lyu, M., Lin, Z., Li, G. \& Situ, G. Fast modal decomposition for optical fibers
%using digital holography. Sci. Rep. 7, 6556 (2017).
%20. Forbes, A., Dudley, A. \& Mclaren, M. Creation and detection of optical modes
%with spatial light modulators. Adv. Opt. Photonics 8, 200 (2016).
%21. Labroille, G. et al. Efficient and mode selective spatial mode multiplexer based
%on multi-plane light conversion. Opt. Express 22, 15599 (2014).
%22. Fontaine, N. K. et al. Laguerre?Gaussian mode sorter. Nat. Commun. 10, 1865
%(2019).
%%%%%%%%%%
%24. Br\"uning, R., Gelszinnis, P., Schulze, C., Flamm, D. \& Duparr\'e, M.
%Comparative analysis of numerical methods for the mode analysis of laser
%beams. Appl. Opt. 52, 7769?7777 (2013).
%25. L\"u, H., Zhou, P., Wang, X. \& Jiang, Z. Fast and accurate modal decomposition
%of multimode fiber based on stochastic parallel gradient descent algorithm.
%Appl. Opt. 52, 2905?2908 (2013).
On the other hand, a number of methods without a reference beam have been also proposed, see references in \cite{turitsynNC20}.
% to solve the MD problem. 
Here, we use a (non-interferometric) numerical computing mode decomposition procedure that is based on the well-established  Gerchberg-Saxton algorithm \cite{shapira05,shechtman15}.
It allows us to retrieve the transverse phase profile of the field from the NF and the
FF intensity distributions measured in the experiments \cite{fienup78,fienup82}.
By projecting the retrieved spatial profile of the complex field over the fiber modes, we get the complete modal distribution of the experimental optical beam $n_p^{\rm exp}$.
We refer the reader to the Ref.\cite{PRL23} (Supplementary Material) for a detailed presentation of the procedure (theoretical and numerical computation of the error introduced by the Gerchberg-Saxton algorithm, impact of the sampling due to the cameras, validity of the phase reconstruction).
In addition, we show experimentally and theoretically in the Appendix that the modal distribution recorded experimentally $n_p^{\rm exp}$ at the fiber output converges toward the expected RJ distribution as the number of realizations of speckles beams is increased.

\begin{widetext}
\begin{center}
\begin{figure}
\includegraphics[width=.8\columnwidth]{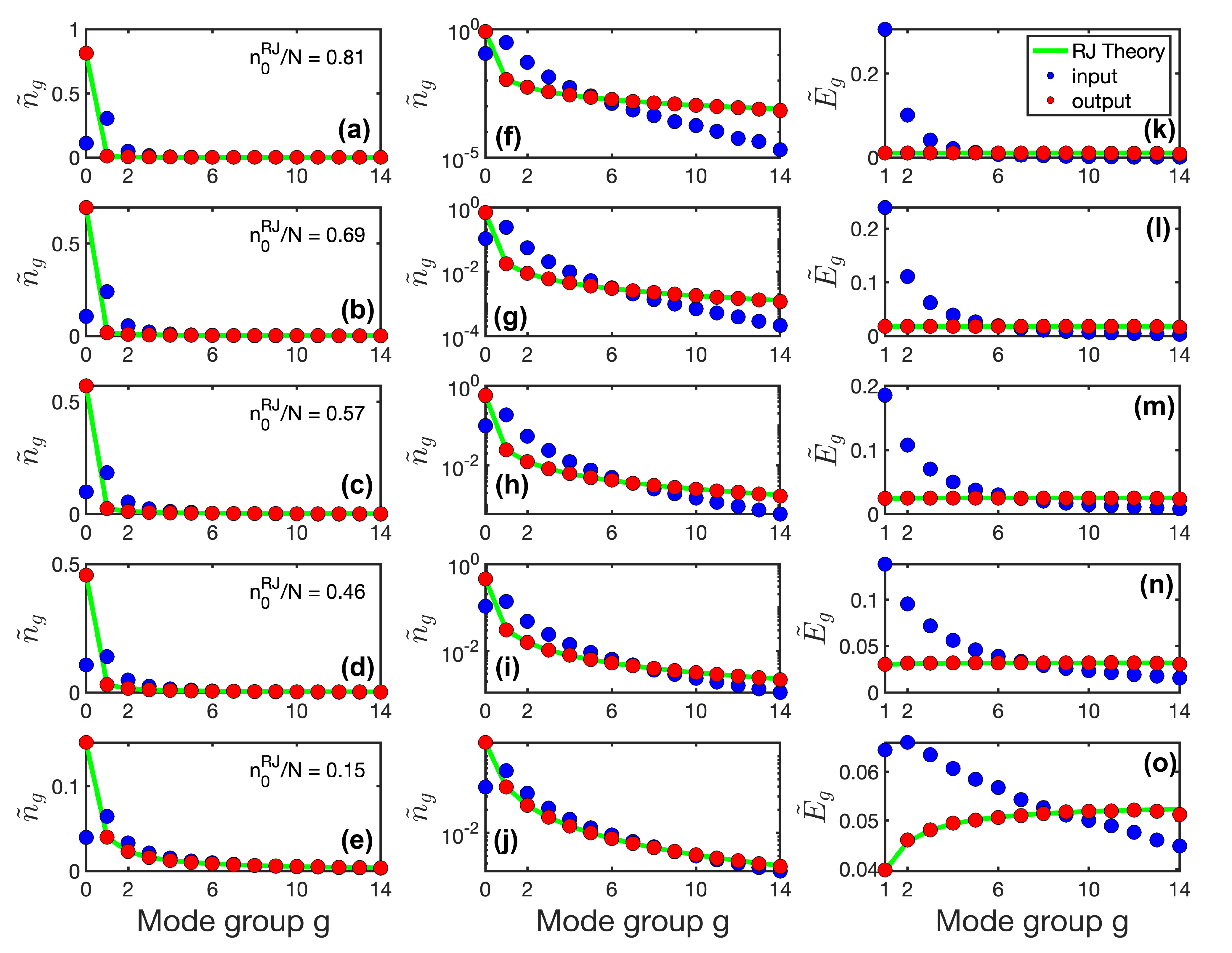}
\caption{
\baselineskip 10pt
{\bf 1st Experiment: Rayleigh-Jeans thermalization.} 
Modal distribution ${\tilde n}_g$ vs mode group $g$ in normal scale (1st column), in log-scale (2nd column), and corresponding energy distribution ${\tilde E}_g$ vs $g$ (3rd column).
Initial modal distribution at the fiber input (blue), and modal distributions at the fiber output (red).
Corresponding RJ distribution (green line): The quantitative agreement with the output distribution is obtained without using any adjustable parameter.
Different lines in the figure correspond to different energies $E$, i.e., different condensate fractions (see the legend in the 1st column).
}
\label{fig:1} 
\end{figure}
\end{center}
\end{widetext}

\begin{figure}
\includegraphics[width=1\columnwidth]{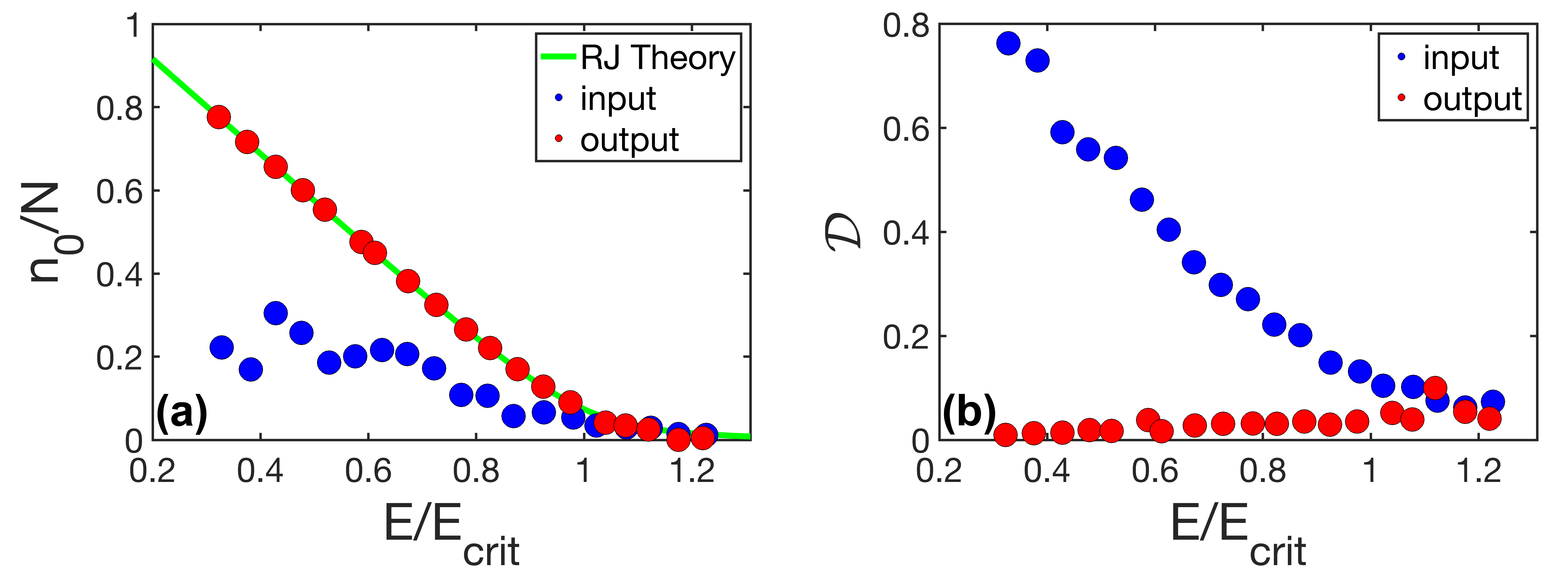}
\caption{
\baselineskip 10pt
{\bf 1st Experiment: Attraction to the RJ distribution.} 
(a) Power fraction into the fundamental mode (condensate fraction) $n_0/N$ vs $E/E_{\rm crit}$: at the fiber input (blue), and fiber output (red).
% Modal distribution $n_g$ vs mode group $g$ (left column), and corresponding energy distribution ${\tilde E}_g$ vs $g$.
%Initial modal distribution at the fiber input (blue), and modal distributions at the fiber output (red).
Corresponding RJ theory (green line): The quantitative agreement with the output distribution is obtained without using any adjustable parameter.
(b) Distance ${\cal D}$ vs energy [defined from Eq.(\ref{eq:distance})] to the RJ equilibrium distribution at the fiber input (blue), and fiber output (red).
The strong reduction of ${\cal D}$ from the input-to-output shows the attraction to the RJ distribution. 
Note that ${\cal D}$ is normalized, $0 \le {\cal D} \le 1$.
}
%\label{fig:2} 
\end{figure}

\subsection{RJ thermalization over a broad energy range}
\label{subsec:RJ_therm_en_rge}

The experiment is realized in the weakly nonlinear regime, where linear propagation effects dominate over nonlinear effects
$L_{lin} \sim \beta_0^{-1} \ll L_{nl}=1/(\gamma N)$,
where $\gamma$ is the nonlinear coefficient of the MMF.
Typically, we have $L_{lin} \sim 0.2$mm, and $L_{nl}\sim 0.3{\rm m}$.
%so that the focusing nonlinearity inherent to optical fibres (i.e., positive nonlinear coefficient $\gamma$) does not play any role in the experimental results.
%Indeed, the speckled pattern propagating through the MMF fluctuates over a linear propagation length $L_{lin}$ much smaller than the nonlinear length $L_{nl}$:
%The speckle beam that propagates through the MMF exhibits fluctuations that vary over a linear propagation length $L_{lin}$ much smaller than the nonlinear length $L_{nl}$:
%The speckle beam that propagates through the MMF exhibits fluctuations that vary over a linear propagation length $L_{lin}$ much smaller than the nonlinear length $L_{nl}$:
%\begin{eqnarray}
%L_{lin} \sim \beta_0^{-1} \sim 0.2{\rm mm} \ \ll \ L_{nl}=1/(\gamma N) \sim 0.3{\rm m},
%\label{eq:Llin_Lnl}
%\end{eqnarray}
This means that the nonlinearity plays a role over a propagation length much larger than that of the rapidly fluctuating speckle beam.

The process of RJ thermalization in the MMF is driven by the Kerr nonlinearity, as described theoretically by a wave turbulence kinetic approach \cite{PRA11b,PRL19,PRA19,PRL22}.
Accordingly, as the beam propagates through the MMF, it is expected to relax toward the thermodynamic equilibrium state described by the RJ distribution \cite{PRA11b,PRL19,PRA19,christodoulides19,PRL20,EPL21,wise_NP22,mangini22}
\begin{equation}
n_p^{\rm RJ}=T/(\beta_p-\mu),
\label{eq:rj}
\end{equation}
where $T$ and $\mu$ are the temperature and chemical potential. 
Accordingly, we have at equilibrium $N=T\sum_p (\beta_p-\mu)^{-1}$ and $E=T \sum_p \beta_p /(\beta_p-\mu)$. 
Important to notice, $(T,\mu)$ are uniquely determined by $(N,E)$, i.e., there is a one to one correspondence between $(T,\mu)$ and $(N,E)$.
In other terms, we are dealing with the microcanonic statistical description, so that $T$ does not refer to the `room' temperature ($T$ is in units of W.m$^{-1}$).
Also note that the RJ distribution can be generalized by taking into account for the conservation of linear momentum \cite{PRL06,EPL07,OE09,PRA09} and angular momentum \cite{pod22,fan22}.
However, the conservation of angular momentum is relevant when peculiar conditions of injection of the optical beam into the MMF are considered, see Ref.\cite{pod22}. 
Here, we consider speckle beams for which the angular momentum 
may vary from one realization to another and vanishes in average, so that its conservation is not relevant to our experiments.

%: The solutions to these equations show that $(\mu,T)$ are uniquely determined by $(N,E)$ \cite{christodoulides19}.
%In analogy with the ideal equilibrium Bose-Einstein condensation, 

At variance with previous experiments of spatial beam cleaning and RJ thermalization \cite{krupa16,wright16,krupa17,PRL20,fabert20,EPL21,wise_NP22}, here we study the process of thermalization for different amount of the (kinetic) energy $E$.
Indeed, by passing the laser beam through a diffuser before injection in the MMF, we can vary the amount of randomness (i.e., degree of spatial coherence) of the speckle beam by keeping fixed the power $N$ -- the higher the randomness of the fluctuations in the speckle beam, the higher the corresponding energy $E$.
In other terms, owing to the diffuser, we  study the process of RJ thermalization over a broad range of energies $E$.
This is at variance with previous experiments \cite{krupa16,wright16,krupa17,PRL20,fabert20,EPL21,wise_NP22}, as will be discussed in the framework of the 2nd experiment presented below in the paper.
Note that throughout the paper, the terminology `coherence' refers to the purely spatial coherence properties of the quasi-monochromatic optical beam that propagates through the fiber.

An other important advantage of using a diffuser is that it allows us to perform an average over an ensemble of realizations.
Indeed, the RJ distribution is in essence a statistical distribution, so that the comparison with the experimental data requires an average over realizations.
We have recorded 2$\times$1000 realizations of the NF and FF intensity distributions for the same power $N$  ($N=7$kW) and different values of the energy $E$. 
For each individual realization, we have retrieved the modal distribution $|a_p^{{\rm exp}}|^2$.
We have  partitioned such an ensemble of 1000 realizations of $\{|a_p^{{\rm exp}}|^2\}$ within small intervals of energies $\Delta E =0.125 E_0$, where $E_0=N \beta_0$ is the minimum of the energy 
%(all of the power populate the fundamental mode).
(only the fundamental mode is populated).
Next, we have performed an average over the realizations of the modal distributions for each specific energy interval $\Delta E$, which gives the averaged modal distributions ${n_p^{\rm exp}}$ for different energies $E$. % vs energy $E$.
We have followed this experimental procedure at the output of the MMF (at $L=12$m), and then we have repeated the procedure at the fiber input (after 20cm of propagation).
%We have  partitioned such 2$\times$1000 realizations within intervals of energies $\Delta E= 0.05 E_{\rm crit}$, i.e., the $m-$th interval being $\Delta E_{\rm exp}^{(m)} = [m \Delta E, (m+1) \Delta E[$ and $E^{(m)}_{\rm exp}$ denotes the averaged energy in this interval.
%We performed an average over the realizations of the modal distributions for a specific energy interval $\Delta E_{\rm exp}^{(m)}$, which gives the averaged modal distributions ${\overline n_p^{\rm exp}}$.

\subsubsection{Power distribution among the modes}

We report in Fig.~2 (left column) the averaged modal distribution in the experiment ${n_p^{\rm exp}} $ at the fiber input and the fiber output, for different energies $E$.
For convenience we have reported the average power ${\tilde n}_g$ within each group of degenerate modes, where $g=0,\ldots,g_{max}-1$ indexes the mode group.
%In the same way, we also report in Fig.~1 (right column) the average energy within each group of degenerate modes, ${\tilde E}_g=\beta_0 g {\tilde n}_g$. 
For the MMF used in the experiments, $g_{max}=15$ for $M=g_{max}(g_{max}+1)/2=120$ modes.

We compare the experimental data with the theoretical RJ distribution $n_p^{\rm RJ}$.
Note that there are no adjustable parameters between the experimental distribution ${n_p^{\rm exp}}$ and the RJ distribution $n_p^{\rm RJ}$: The parameters $(T,\mu)$ in $n_p^{\rm RJ}$ in Eq.(\ref{eq:rj}) are uniquely determined by the power and the energy $(N,E)$ measured in the experiments.
We observe in Fig.~2 a very good agreement between the observed modal populations ${n_p^{\rm exp}}$ and the RJ distribution $n_p^{\rm RJ}$.
Furthermore, such a good agreement is obtained over a broad range of the energy $E$.

In order to properly appreciate the key role of the energy $E$, we report in Fig.~3(a) the condensate fraction (i.e., amount of power fraction in the fundamental fiber mode) $n_0^{\rm RJ}/N$ vs $E/E_{\rm crit}$, where $E_{\rm crit} \simeq N \beta_0 \sqrt{M/2}$ is the threshold value of the energy below which the fundamental mode gets macroscopically populated \cite{PRL20}. 
In a loose sense, the energy $E$ plays a role analogous to the temperature, i.e., the condensate fraction $n_0^{\rm RJ}/N$ increases as the energy $E$ decreases.
Note in particular that the effect of beam self-cleaning is suppressed nearby the transition to condensation for $E \simeq E_{\rm crit}$, a feature discussed in detail in Ref.\cite{PRL20}.

\subsubsection{Energy distribution among the modes}

The modal distribution of the power is mainly sensitive to the behaviour of the low-order modes.
So as to properly characterize the thermalization process over all modes, it proves convenient to consider the distribution of the kinetic energy $E_p=(\beta_p-\beta_0) n_p$, or equivalently the average energy within groups of degenerate modes ${\tilde E}_g=\beta_0 g {\tilde n}_g$, where we recall that $g$ indexes the mode group.
The results reported in Fig.~2 (right column) confirm a remarkable quantitative agreement between the experimental data and the RJ equilibrium distribution.
We note in particular that for a small value of the energy $E \ll E_{\rm crit}$, we can observe an {\it energy equipartition among the modes}, a property inherent to {\it classical statistical mechanics}.
Indeed, well below the energy threshold for condensation, i.e. $E \ll E_{\rm crit}$, the chemical potential reaches the fundamental mode eigenvalue $\mu \to \beta_0^-$, so that $n_p^{\rm RJ}=T/(\beta_p-\beta_0)$ for $p \neq 0$ (or ${\tilde n}_g=T/(g \beta_0)$ for $g\neq 0$), and thus 
\begin{equation}
E_p = T \quad {\rm for} \quad p \neq 0 
\end{equation}
(or ${\tilde E}_g=T$ for $g \neq 0$). 
This property of energy equipartition can be observed in all panels (k-n), ${\tilde E}_g^{\rm exp}=T$, the only panel which shows a significant departure from energy equipartition is panel (o) because $E$ is then close to $E_{\rm crit}$.
Note that this is consistent with the fact that for $E \gtrsim E_{\rm crit}$, the chemical potential deviates from the fundamental mode eigenvalue, and the RJ distribution (\ref{eq:rj}) with $\mu \ll \beta_0$ no longer verifies the property of energy equipartition among the modes.

The experimental results also evidence the double turbulence cascade underlying RJ thermalization: 
While the power flows toward the fundamental mode (inverse cascade), the flow of kinetic energy toward the higher-order modes (direct cascade).
This latter effect is shown in panels (l-o) of Fig.~2, for a relative small condensate fraction $n_0^{\rm RJ}/N =0.15$.
This double cascade process was already commented in Ref.\cite{EPL21} through the analysis of the experimental data in direct $\br-$space.
The mode-resolved analysis reported in Fig.~2 is more appropriate and convincing. 
%For this purpose, we introduce the modal energy $E_p=(\beta_p - \beta_0) n_p^{\rm exp}$.

\subsubsection{Attraction to the RJ distribution}

We can estimate quantitatively the attraction to the RJ equilibrium by defining a `distance' to the RJ distribution.
For this purpose, we define a quantity that reflects the degree of similarity with the RJ distribution in mode space:
\begin{equation}
{\cal D} = \frac{\sum_p |n_p^{\rm exp} - n_p^{\rm RJ}|}{ \sum_p n_p^{\rm exp} + n_p^{\rm RJ}}.
\label{eq:distance}
\end{equation}
Note in particular that this quantity is normalized, $0 \le {\cal D} \le 1$.
We report in Fig.~3(b) the distance ${\cal D}$ computed for the experimental data $n_p^{\rm exp}$ at the fiber input (blue), and at the fiber output (red), for different energies $E$.
The strong reduction of the distance ${\cal D}$ from the fiber input to the fiber output clearly confirms the process of thermalization to the RJ equilibrium.
It is important to note that this attraction process takes place over a broad range of variation of the energy $E$. 
Note that, as the energy $E$ increases and approaches the critical value of the transition to condensation $E_{\rm crit}$, the input-to-output reduction of ${\cal D}$ decreases as well, a feature that becomes apparent in Fig.~2 (left column), where we can notice that the input and output distributions become similar as the energy increases.

%The distance ${\cal D}_N$ is quite sensitive to the behaviour of the low-order modes.
%In order to properly characterize the thermalization process over all of the modes of the MMF, it proves convenient to introduce an analogous distance for the energy:  
%\begin{equation}
%{\cal D}_E = (M-1) \frac{\sum_{p\neq 0} ({\cal E}_p^{\rm exp}-{\cal E}_p^{\rm RJ})^2}{\big[ \sum_{p\neq 0} ({\cal E}_p^{\rm exp}-{\cal E}_p^{\rm RJ}) \big]^2},
%\end{equation}
%where ${\cal E}_p^{\rm exp}=(\beta_p - \beta_0) n_p^{\rm exp}$ and ${\cal E}_p^{\rm RJ}=(\beta_p - \beta_0) n_p^{\rm RJ}$.
%Again, the significant reduction of the distance ${\cal D}_E$ from input-to-output beams clearly strengthens the process of thermalization to the RJ equilibrium.

%Leaky modes?:
%The MMF has a core, a cladding (with radius 62.5$\mu$m), and a highly absorbing polymer-coating with refractive index larger than the core. 
%Then leaky radiation modes in the cladding are rapidly absorbed during propagation due to their large penetration in the polymer-coating: we measured a typical absorption length of $\sim$15cm. 
%Since this length is much smaller than that required to generate leaky modes, the nonlinear excitation of leaky modes can be neglected.

\begin{figure}
\includegraphics[width=1\columnwidth]{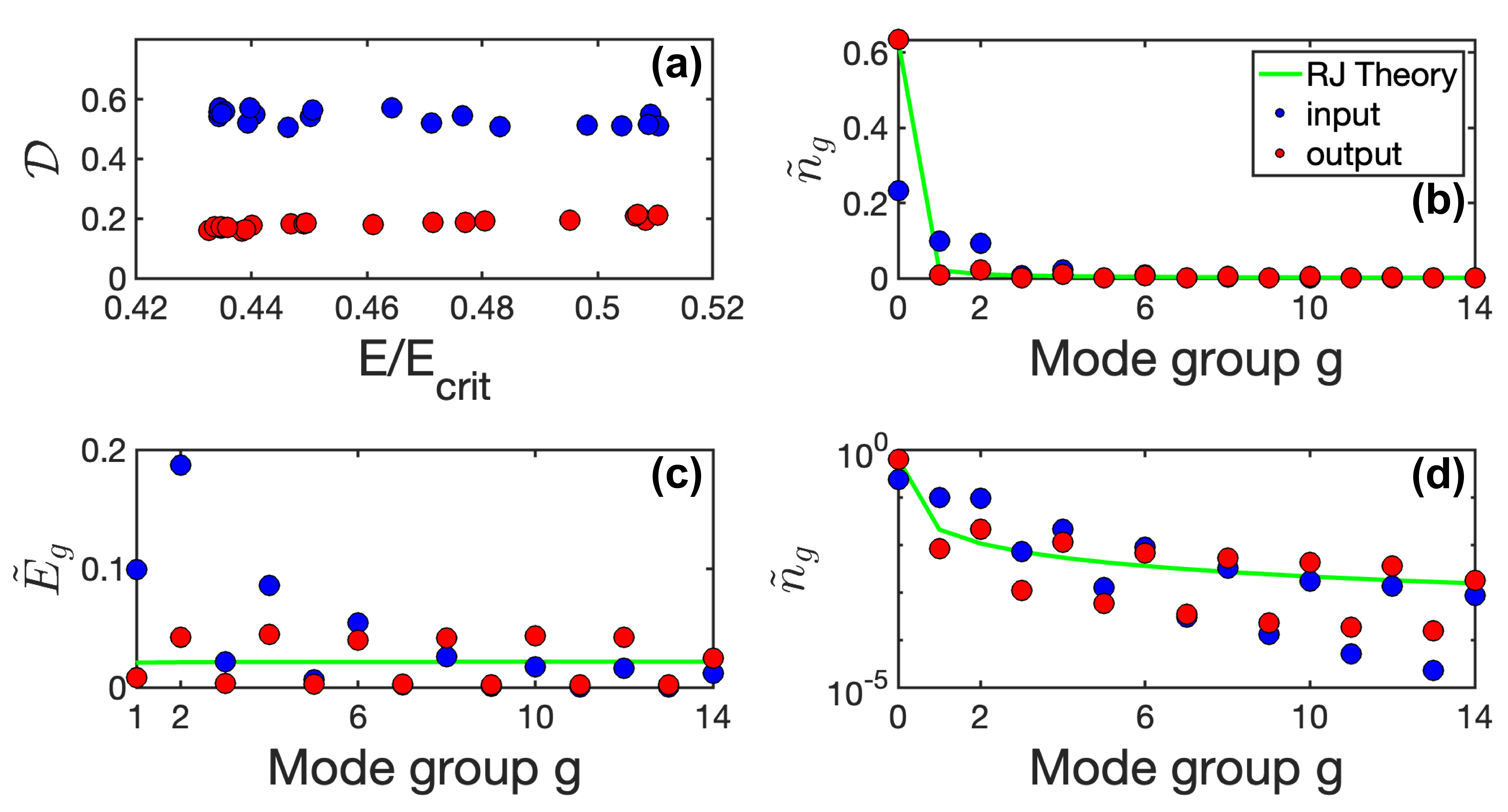}
\caption{
\baselineskip 10pt
{\bf 2nd Experiment: Incipient RJ thermalization.} 
(a) Distance ${\cal D}$ vs energy $E/E_{\rm crit}$ [defined from Eq.(\ref{eq:distance})] to the RJ equilibrium distribution at the fiber input (blue), and fiber output (red).
(b)-(d) Modal distribution ${\tilde n}_g$ vs mode group $g$ (right column), and corresponding energy distribution ${\tilde E}_g$ vs $g$.
Initial modal distribution at the fiber input (blue), and modal distributions at the fiber output (red), and corresponding RJ distribution (green line).
Note that ${\cal D}$ in (a) is bounded, $0 \le {\cal D} \le 1$.
}
%\label{fig:2} 
\end{figure}

\section{Second experiment: Beam cleaning through incipient thermalization}

\subsection{Differences with the first experiment}

There are two differences that distinguish the 1st experimental configuration with the 2nd configuration.
(i) In the 2nd experiment the MMF length is reduced by a factor 4 to $L=3$m -- accordingly  the power is increased by the same factor to $N=28$kW, so as to maintain constant the effective number of nonlinear lengths $L/L_{nl}$ in both experiments. 
Because of the short fiber length considered, the impact of disorder (weak random coupling) is significantly reduced in this 2nd configuration.
(ii) We do not pass the laser beam through a diffuser before injection into the MMF, i.e., we launch a Gaussian-shaped {\it coherent beam} (at approximate normal incidence) into the fiber.
Notice that the absence of the diffuser in this 2nd configuration does not allow us to make an average over the realizations.
The energy $E$ is varied by varying the radius of the launched Gaussian beam: For a given power $N$, by increasing the radius of the Gaussian beam, the population of modes at the fiber input is increased, which in turn increases the energy $E$.

\subsection{Deviation from the RJ distribution}

Following a procedure similar to that carried out in the 1st experiment, we report in Fig.~4 the modal distribution of the power 
%$n_p^{\rm exp}$,
$\tilde{n}_g^{\rm exp}$, 
and the energy ${\tilde E}_g^{\rm exp}$, observed in the 2nd experiment, at the fiber input (blue), and the fiber output (red).
We compare the experimental data with the theoretical RJ distribution (green line).
We can note that the low-order modes are efficiently attracted toward the RJ equilibrium distribution, while the higher-order modes exhibit some significant deviation to the RJ equilibrium.
This aspect becomes even more apparent through the analysis of the distribution of the kinetic energy 
%$E_p^{\rm exp}$ 
$\tilde{E}_g^{\rm exp}$
reported in Fig.~4(c).
We can remark that mode groups containing the radially symmetric modes (with $g$ even) are more populated than the other (with $g$ odd), as a natural consequence of the injection of a radially symmetric coherent beam in the MMF.
Aside from symmetry considerations, note that nonlinear modal interactions can have different efficiencies also due to the large variation in nonlinear overlap coefficients among the modes, which can significantly affect the rate of convergence to equilibrium \cite{PRA19}.
Finally note that the deviation from the RJ distribution is also evidenced by the computation of the distance ${\cal D}$. 
%\textcolor{red}{Est-ce ${\cal D}$ ou bien est-ce
%\begin{equation}
%\tilde{\cal D} = \frac{\sum_g |\tilde{n}_g^{\rm exp} - \tilde{n}_g^{\rm RJ}|}{ \sum_g %\tilde{n}_g^{\rm exp} + \tilde{n}_g^{\rm RJ}} ~~?
%\label{eq:tildedistance}
%\end{equation}
%}
As illustrated in Fig.~4(a), the input-to-output decrease of ${\cal D}$ shows that the system approaches the RJ distribution, although a significant deviation from that distribution is visible at the fiber output.

\begin{figure}
\includegraphics[width=1\columnwidth]{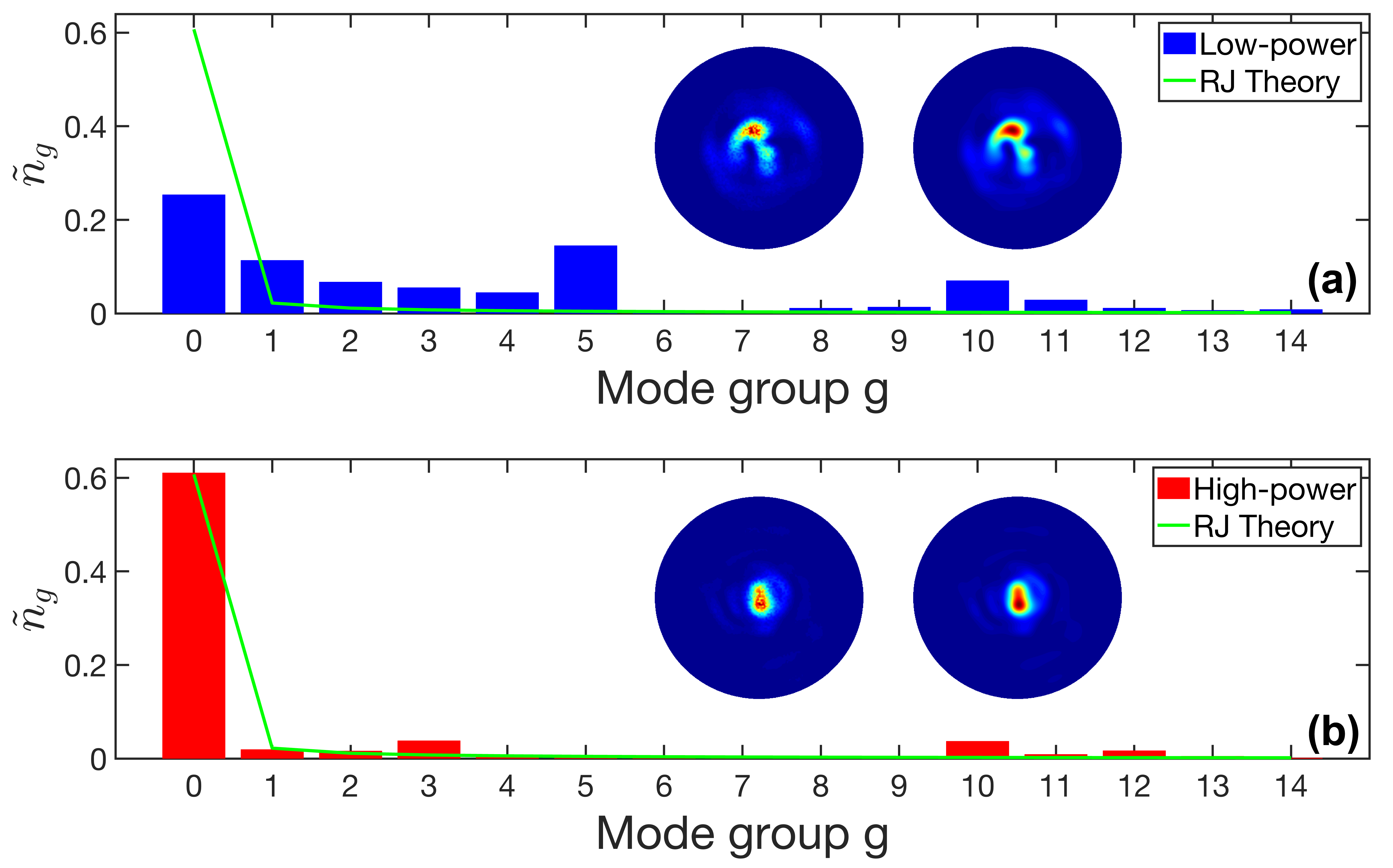}
\caption{
\baselineskip 10pt
{\bf 2nd Experiment: Beam self-cleaning.} 
Modal distribution ${\tilde n}_g$ vs mode group $g$, in the presence of a small power $N=0.23$kW (linear regime) (a), and high power $N=28$kW (nonlinear regime) (b).
The green line reports the RJ equilibrium distribution.
The insets show the 2D intensity distributions: measured (left) and reconstructed (right) output beam profile.
}
%\label{fig:2} 
\end{figure}

%Consider for instance the case where a Gaussian beam with a radius comparable to that of the fundamental mode is injected at perfect normal incidence and exactly at the center of the MMF. In this case only few radial modes are excited, while all modes featured by a nonhomogeneous azimuthal profile are not excited at all.

Although the 2nd experimental configuration only shows an incipient process of RJ thermalization, it evidences a process of beam self-cleaning.
This is illustrated in Fig.~5, which reports the modal distribution
% $n_p^{\rm exp}$ 
$\tilde{n}_g^{\rm exp}$ at low-power, and at high-power, at the output of the MMF.
The corresponding 2D intensity distributions at small and high power  shows an effect of spatial beam cleaning that is dominated by the behavior of the low order modes, although the higher-order modes significantly deviate from the RJ equilibrium distribution.

\section{Interpretation and conclusion}

We have studied the impact of disorder on the process of thermalization in multimode optical fibers by comparing two different experiments realized with a long fiber and an incoherent excitation (1st experiment), and a short fiber with a coherent excitation (2nd experiment).
The results of the 1st experiment demonstrate without ambiguity the process of thermalization to the RJ equilibrium distribution, over a broad range of the energy $E$.
For small values of $E$, the strong condensate fraction ($n_0 \gg n_{p \neq 0}$) entails a process of beam-cleaning during the propagation through the MMF.
In the 2nd experiment, an effect of beam-cleaning can be identified, which can be ascribed to the RJ thermalization of low-order modes, while high-order modes can exhibit a deviation from the RJ distribution.

\begin{figure}
\includegraphics[width=1\columnwidth]{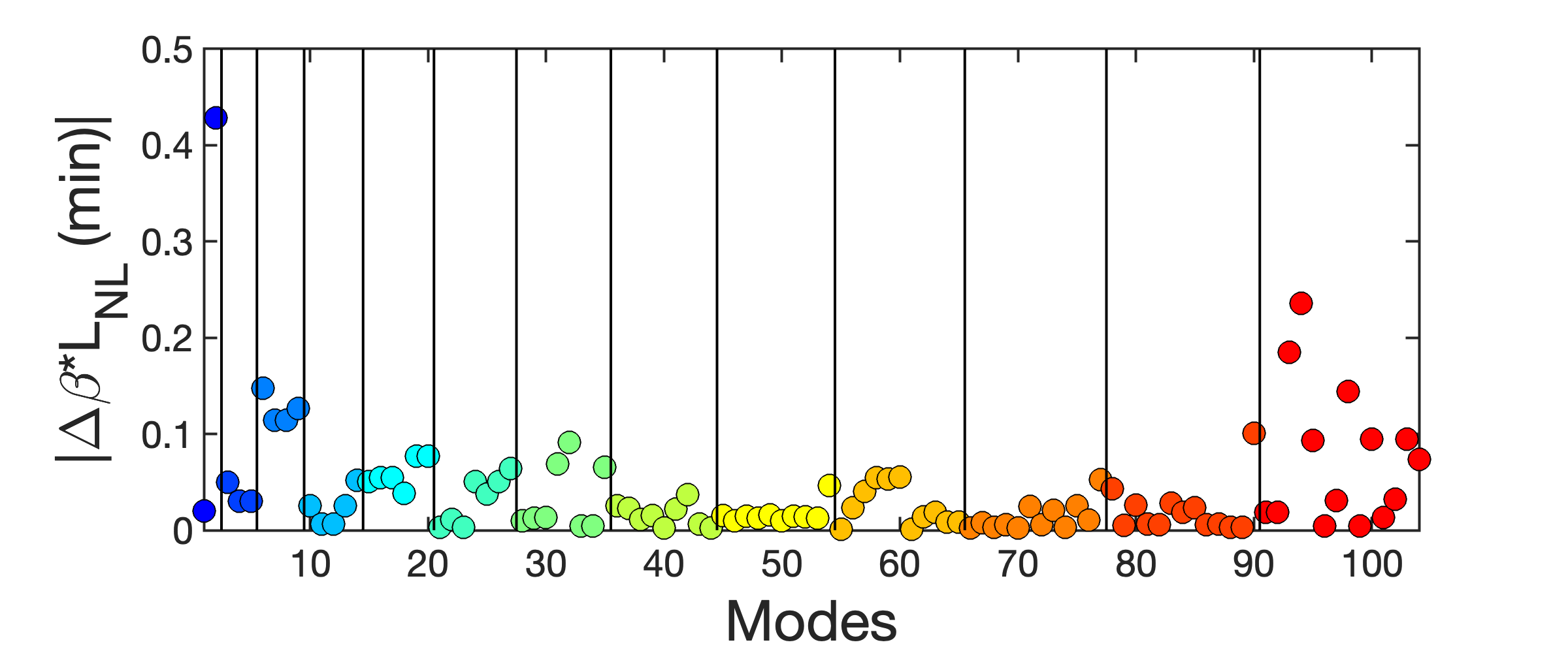}
\caption{
\baselineskip 10pt
{\bf Quasi-resonances accounting for the truncated potential.} 
Computation of the most favourable non-trivial resonances involving three groups of modes $\Delta \beta_g = {\rm min}(2 \beta_g -\beta_{g-1}-\beta_{g+1})$, and corresponding resonance efficiency $|\Delta \beta_g| L_{nl}$, where $\beta_g$ are computed by taking into account the truncation of the parabolic potential due to the fiber cladding.
%with the nonlinear length of our experiments $L_{nl}=0.3$m.
The modes of the fiber can find efficient quasi-resonances $|\Delta \beta_g| L_{nl} \lesssim 1$ that contribute to the thermalization process observed in the 1st experiment, see Fig.~2.
}
%\label{fig:2} 
\end{figure}

\subsection{Discussion of previous results}

Previous experimental results  
%a detailed comparison of the measured modal distribution 
revealed a good agreement with the theoretical RJ distribution, especially for low-order modes \cite{wise_NP22,mangini22}.
%In the recent work \cite{mangini22}, a direct comparison has been reported for both low-order modes and high-order modes.
On the other hand, some deviation between the experimental measurements and the theoretical RJ distribution was evidenced for high-order modes when the spatial azimuthal complexity grows, see Fig.~8 in Ref.\cite{mangini22}.
The results in Ref.\cite{mangini22} were obtained with a short fiber length and the injection of coherent laser beams populating radially symmetric modes, a feature which seems to corroborate the results of our 2nd experiment.
%: A process of beam-cleaning condensation can be identified owing to the thermalization of low-order modes, while high-order modes exhibit some deviation from the RJ distribution.
%This is at variance with our 1st experiment, which evidences a process of RJ thermalization over all fiber modes, and over a broad range of the energy $E$.

%It turns out that the modes which have radial symmetry are the most populated in the occurrence of BSC. Whereas, the mode occupancy vanishes when the spatial (azimuthal) complexity grows.
 
A good agreement has been also obtained between the measured modal distribution and the theoretical RJ distribution for low-order groups of fiber modes \cite{wise_NP22}.
The discrepancy between the experiments and the RJ theory for high-order modes was attributed to the lack of efficient resonances among the modes.
Indeed, in an ideal parabolic potential, the regular spacing of the mode groups provides efficient phase-matching resonances. 
In practice however the potential is truncated by the presence of the cladding, which distorts the ladder of mode-group eigenvalues, especially for higher-order modes that overlap more strongly with the cladding.
We have computed numerically the mode eigenvalues accounting for the truncation of the parabolic potential for the MMF used in our experiments.
We have considered the efficiency of the most favourable non-trivial resonances involving three groups of modes $\Delta \beta_g = {\rm min}(2 \beta_g -\beta_{g-1}-\beta_{g+1})$, where we recall that $g$ indexes the mode group.
Then we compute the corresponding resonance efficiency $|\Delta \beta_g| L_{nl}$, where $L_{nl}$ denotes the nonlinear length in the 1st experiment $L_{nl} \simeq 0.3$m.
Only quasi-resonances verifying $|\Delta \beta_g| L_{nl} \lesssim 1$ contribute to the thermalization process, while four-mode interactions such that $|\Delta \beta_g| L_{nl} \gg 1$ are non-resonant and do not contribute. 
We report in Fig.~6 the most favourable non-trivial resonances for each fiber mode, with  $g=1,...,13$ (remember $g_{max}=15$).
We can see in Fig.~6 that even higher-order modes of the fiber can find an efficient non-trivial quasi-resonance verifying $|\Delta \beta_g| L_{nl} \lesssim 1$, which indicates that the finite truncation of the parabolic potential should not prevent the thermalization of high-order modes.
This corroborates the results reported in the 1st experiment where RJ thermalization is observed over all groups of fiber modes. 
We also note that the validity of the results reported throughout this paper are not affected by the correction due to the finite truncation of the parabolic potential, i.e., plotting Figs.~1-4 with the correction does not provide visible differences in the figures.

%\begin{widetext}
%\begin{center}
\begin{figure}
\includegraphics[width=1\columnwidth]{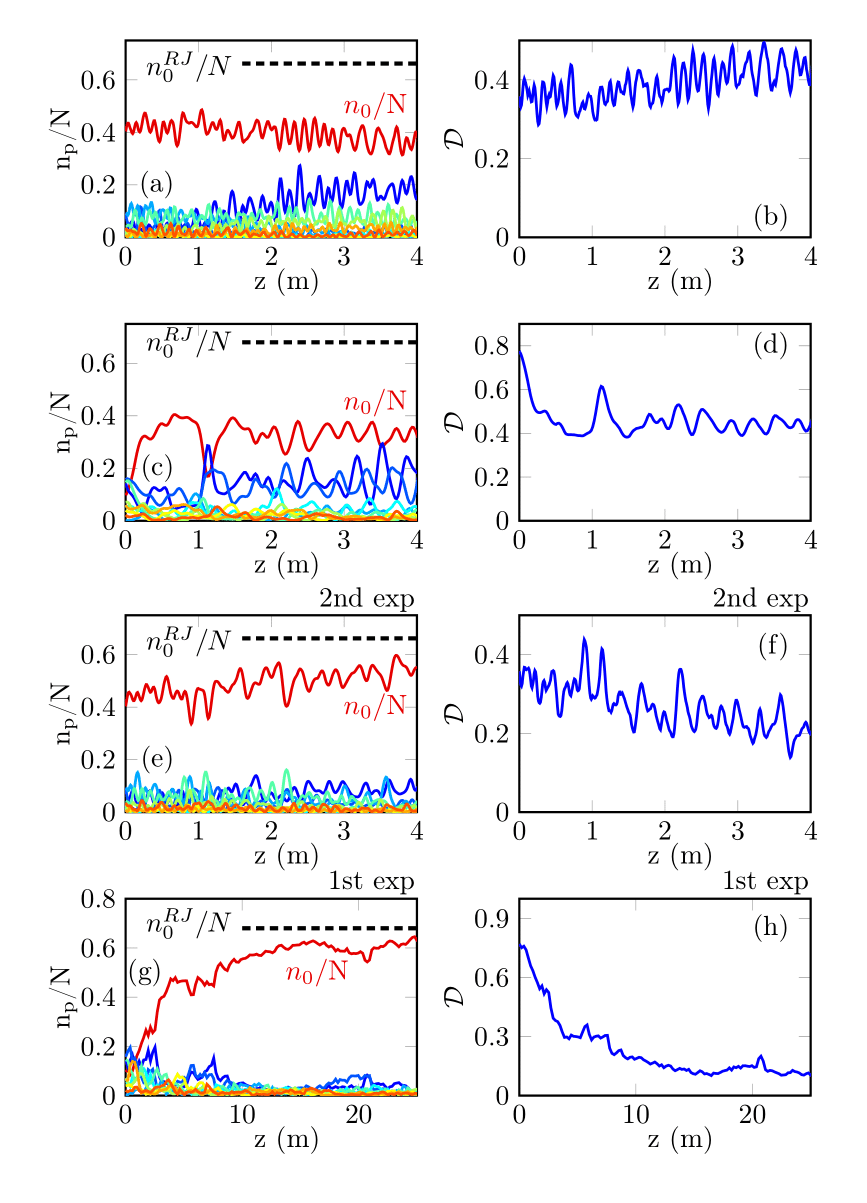}
\caption{
\baselineskip 10pt
{\bf Numerical simulations:} 
Simulations of the NLS equation showing the evolutions of the modal
components $n_p(z)/N$ (left column), and corresponding evolution of the distance ${\cal D}(z)$ to the RJ distribution (right column).
Fundamental mode $p = 0$ (red), $p = 1$ (dark blue), $p = 2$ (blue), $p = 3$ (light blue), $p = 4$ (cyan).
% $p = 5$ (light green), $p = 6$ (green), $p = 7$ (yellow), and $p = 8$ (orange). 
1st row (a)-(b): Simulation without disorder (weak random mode coupling) and starting from a coherent initial condition. 
2nd row (c)-(d): Same simulation as (a)-(b), except that the initial condition is a speckle beam.
In (a-d), the system exhibits a coherent regime of mode interaction featured by persistent oscillations that prevent RJ thermalization. 
3rd row (e)-(f): Same simulation as (a)-(b), except that disorder is included during the propagation, which corresponds to the 2nd experimental configuration. 
4th row (g)-(h): Same simulation as (e)-(f), except that the initial condition is a speckle beam and that the power has been reduced by a factor 4, i.e., 12m of propagation in the bottom row corresponds to 3m in 3rd row -- the bottom row corresponds to the 1st experimental configuration. 
The impact of disorder is more effective in the bottom row (1st experimental configuration)  as compared to the 3rd row (2nd experimental configuration): Disorder accelerates RJ thermalization to equilibrium, as revealed by the evolutions of ${\cal D}(z)$.
The horizontal dashed lines (left column) denote the condensate fractions at RJ equilibrium.
%The insets show the intensity distributions into the fiber core.
See the text for parameters.}
%\label{fig:1} 
\end{figure}
%\end{center}
%\end{widetext}

\subsection{Interpretation and simulations}

The substantial differences that distinguish the results of the 1st and  2nd experiments reported in this work may be interpreted along the lines of the theoretical works in Refs.\cite{PRL19,PRA19}.
We illustrate this aspect in Fig.~7, which reports the numerical simulations of the spatial NLS equation in different configurations.
It was shown in Refs.\cite{PRL19,PRA19} that, by ignoring the impact of disorder (weak random mode coupling due to polarization fluctuations), 
%and considering a coherent excitation (i.e., by launching a coherent beam), 
a strong correlation among the modes is preserved for large propagation lengths, which leads to a phase-sensitive ``coherent regime" of modal interaction. 
In this coherent regime, the modes experience a quasi-reversible exchange of power with each other, 
%thus leading to an oscillatory dynamics of the intensity pattern. 
%Such a multimode beam does not exhibit an enhanced brightness that characterizes a stable self-cleaning effect. 
which tends to freeze the process of RJ thermalization.
This regime is illustrated in Fig.~7 (top row), which reports the evolution of the modal components without disorder 
%in the 2nd experimental configuration, i.e., 
with a coherent initial condition.
Note that a similar regime featured by a phase-sensitive ``coherent" modal interaction is also observed by starting the simulation with a speckle beam, see the second row in Fig.~7.
The oscillatory behavior of the modal components and the slowing down of RJ thermalization can be related to the existence of Fermi-Pasta-Ulam recurrences, as recently discussed in Ref.\cite{malomed21,malomed23} in the framework of the weakly nonlinear regime of the 2D NLS equation with a parabolic trapping potential.
%In the weakly nonlinear regime, an effective equation is derived in \cite{malomed21} with additional invariants, which allows to identify Fermi-Pasta-Ulam recurrence in the dynamics.
On the other hand, we recall that a {\it key ingredient for the occurrence of RJ thermalization} is the existence of a random phase dynamics among the  modes \cite{nazarenko11,chibbaro17} (see also \cite{PRX17,onorato20}).
In other words, {\it a mechanism that breaks the coherent phase-dynamics among the modes is required in order to achieve thermalization in the weakly nonlinear regime considered in the experiments}.
A natural mechanism is provided by the structural disorder, which is inherent to light propagation in MMFs and whose leading order contribution is associated to random polarization fluctuations \cite{mecozzi12a,mecozzi12b,mumtaz13,xiao14}.
This weak random coupling regime is sufficient to introduce a dephasing dynamics among the modes.
This is illustrated in the 3rd row of Fig.~7, that reports the same simulation (same coherent initial condition and same parameters) as in the top row, except that a weak random mode coupling has been introduced, with a correlation length $\ell_c =0.3$m, and a strength of disorder $2\pi/\sigma=2.1$m, see Ref.\cite{PRL19,PRA19} for details.
The simulation in the 3rd row of Fig.~7 then corresponds to the 2nd experimental configuration.
We note in Fig.~7 that the oscillatory behavior of the modal components associated to the coherent modal phase dynamics is less pronounced in the 3rd row than in the 1st or 2nd rows:
The evolution of the population of the fundamental mode $n_0(z)$ keeps an oscillating behavior, although such a behavior tends to slowly approach the RJ equilibrium value (horizontal dashed  black line, $n_0^{\rm RJ}/N \simeq 0.65$).
This observation is corroborated by the evolution of the distance ${\cal D}(z)$ to the RJ equilibrium that also keeps an oscillatory behavior, though it tends to lower during the propagation (panel (f) of Fig.~7), a feature which is in contrast with the corresponding evolution without disorder (panels (b), or (d) for $z>1$m, in Fig.~7).

This observation is confirmed by the numerical simulation realized in the framework of the 1st experimental configuration, see bottom row in Fig.~7 with an injected speckle beam.
In this case a larger fiber length is considered by decreasing the power by a factor 4, so as to keep constant the effective number of nonlinear interaction lengths, i.e., 12m of propagation in the bottom row of Fig.~7 corresponds to 3m in the 3rd row 
(or in the 1st and 2nd rows).
Because of the larger fiber length, the impact of disorder is more effective for the 1st experimental configuration (bottom row) as compared to the 2nd configuration (3rd row):
The condensate fraction in Fig.~7(g) reaches the RJ thermal value, while the distance to RJ equilibrium tends to decrease to zero in an almost monotonic fashion, see Fig.~7(h).
Interestingly, the evolutions of $n_0(z)$ and ${\cal D}(z)$ in the 1st experimental configuration (bottom row) do not exhibit the oscillatory behaviour evidenced in the 
2nd configuration (3rd row), which confirms the absence of modal phase-correlations and the significant acceleration of RJ thermalization due to the disorder. 

The simulations reported in Fig.~7 qualitatively reproduce the experimental results. However, it should be stressed that a power larger by a factor three has been considered to accelerate the dynamics in both the 1st and 2nd configurations. 
Then although the purely spatial model considered in the simulations captures the essential features of the experiments, a quantitative description of the experimental results would require a detailed analysis of the temporal averaging effect inherent to the pulsed regime considered in the experimental measurements \cite{Leventoux21a,Leventoux21b,conforti17}.

We finally note that, on the basis of the wave turbulence theory, we have derived a kinetic equation describing the evolutions of the modal components $n_p(z)$,  which revealed that polarization fluctuations significantly accelerate the process of RJ thermalization, see Refs.\cite{PRL19,PRA19}. 
%disorder introduces an effective dissipation that modifies the regularization of four-mode resonances Ref.\cite{PRL19,PRA19}.
This theoretical approach then explains, at a pure qualitative level, the fast process of RJ thermalization observed in the 1st experiment reported in this work.
We recall that, at variance with the 2nd experiment, the 1st experiment is characterized by two factors: (i) a larger fiber length that enhances the impact of disorder, (ii) the injection of a speckle beam owing to the diffuser.
The combination of these two factors leads to a random phase dynamics among the modes, which favours the process of RJ thermalization observed in the 1st experiment.
Finally, this work contributes to the understanding of the fast process of optical thermalization observed in multimode optical fibers, a feature that may also be important for possible future extensions to the thermalization of multiple beams \cite{Ferraro23a,Ferraro23b}. 
From a broader perspective, this work also contributes to the understanding of the interplay of nonlinearity and disorder \cite{cherroret15,cherroret20,cherroret21,wang20,haldar20}, 
in relation with the paradigm of statistical light-mode dynamics, glassy
behaviors, and complexity science \cite{conti22}.

\section{Acknowledgments} 

The authors are grateful to S. Rica, I. Carusotto and V. Doya for fruitful discussions. Fundings: Centre national de la recherche scientifique (CNRS), Conseil r\'egional de Bourgogne Franche-Comt\'e, iXCore Research Fondation, Agence Nationale de la Recherche (ANR-19-CE46-0007, ANR-15-IDEX-0003, ANR-21-ESRE-0040). Calculations were performed using HPC resources from DNUM CCUB (Centre de Calcul, Universit\'e de Bourgogne).

%\bigskip

\section{Appendix: Convergence to RJ equilibrium with the number of speckle realizations}

In this Appendix, we show that the modal distribution recorded experimentally $n_p^{\rm exp}=\left<|a_p^{{\rm exp}}|^2\right>$ converges toward the expected RJ distribution $n_p^{\rm RJ}$ as the number of realizations of speckles beams involved in the average is increased.
Here, we analyze the results retrieved from the 1st experimental configuration, see Figs.~1-2.
The convergence of $n_p^{\rm exp}$ to the RJ distribution follows qualitatively the behavior expected theoretically. 

\begin{figure}
\includegraphics[width=1\columnwidth]{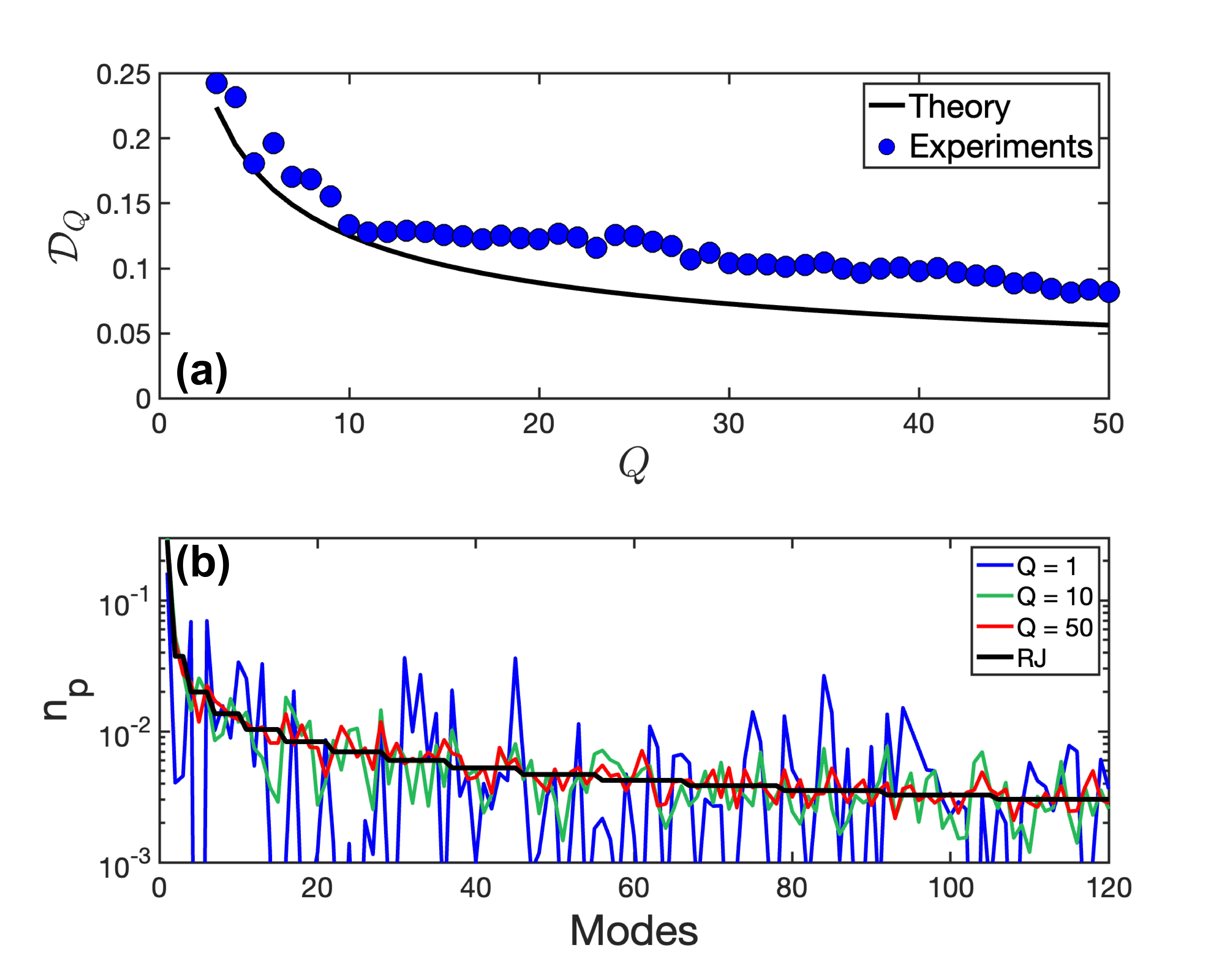}
\caption{
\baselineskip 10pt
{\bf Experimental convergence to the RJ distribution with the number of realizations.} 
(a) The blue points denote the distance ${\cal D}_Q$ [Eq.(\ref{eq:distance_sampl})] between the experimental modal distribution ${n}_p^{{\rm exp},Q}$ averaged over $Q$ realizations of speckle beams, and the theoretical RJ equilibrium distribution ${n}_p^{\rm RJ}$.
The black line reports the theoretical distance ${\cal D}_Q$ computed analytically, from Eq.(\ref{eq:errQtheo}).
Note that ${\cal D}_Q$ is normalized, $0 \le {\cal D}_Q \le 1$.
(b) Experimental modal distribution ${n}_p^{{\rm exp},Q}$ averaged over $Q$ realizations [$Q=1$ (blue), $Q=10$ (green), $Q=50$ (red)].
Corresponding theoretical RJ distribution ${n}_p^{\rm RJ}$ (black line).
}
\label{fig:app_D} 
\end{figure}

As discussed in section~\ref{subsec:RJ_therm_en_rge}, we have recorded in the experiments 2$\times$1000 realizations of the NF and FF intensity distributions for the same power $N=7$kW and different values of the energy $E$. 
For each individual realization, we have retrieved the modal distribution $|a_p^{{\rm exp}}|^2$.
We have partitioned such an ensemble of realizations within small intervals of energies $\Delta E/E_0 =0.125$, with $E_0=N \beta_0$ is the minimum of the energy.
% (all of the power populate the fundamental mode).
Next, we have performed an average over the realizations of the modal distributions for each specific energy interval $\Delta E$.
In the following, we consider a particular energy interval around $E/(N \beta_0)=6$ and we study the convergence to the RJ equilibrium distribution with the number of realizations of speckle beams.
We define the distance between the experimental data and the RJ equilibrium as follows:
\begin{equation}
{\cal D}_{Q} = \frac{ \sum_p \big|{n}_p^{{\rm exp},Q} -
{n}_p^{\rm RJ} \big|}{  \sum_p 
 {n}_p^{{\rm exp},Q} +
{n}_p^{\rm RJ} } .
\label{eq:distance_sampl}
\end{equation}
where the experimental distribution ${n}_p^{{\rm exp},Q} = (1/Q) \sum_{j=1}^Q  |a_p^{{\rm exp},j}|^2$ is the empirical average over $Q$ realizations of the modal distributions $|a_p^{{\rm exp},j}|^2$.
We report the distance ${\cal D}_Q$ vs number of realizations $Q$ in Fig.~\ref{fig:app_D}(a).
% by assuming that the algorithm does not introduce errors.

%\begin{widetext}
%\begin{center}
%\end{center}
%\end{widetext}

We compare the experimental results with the  theoretical expression of the corresponding distance that we have derived in Ref.\cite{PRL23} (Eq.(3) in the Supplementary Methods):
\begin{equation}
{\cal D}_Q \simeq  \frac{Q^{Q-1}}{(Q-1)!} e^{-Q}  .
\label{eq:errQtheo}
\end{equation}
For $Q\geq 8$ , we have ${\cal D}_Q \simeq 1/\sqrt{2\pi Q}$.
Note that Eq.(\ref{eq:errQtheo}) has been found in quantitative agreement with the numerical simulations (see Fig.~4 in the Supplementary Material of Ref.\cite{PRL23}).
We can note that the distance ${\cal D}_Q$ to the RJ distribution computed from the experimental results (blue points) decreases with the number of realizations by following qualitatively the theoretical behavior (solid black line). 
The theoretical curve given by Eq.(\ref{eq:errQtheo}) assumes that the Gerchberg-Saxton algorithm, as well as the sampling of the cameras used in the experiments do not introduce errors. 
The impact of the two errors can explain that the distance ${\cal D}_Q$ for the experimental data reported in Fig.~\ref{fig:app_D}(a) is slightly above the theoretical prediction.
Finally, we report in Fig.~\ref{fig:app_D}(b) the experimental modal distributions ${n}_p^{{\rm exp},Q}$ averaged over $Q$ realizations of speckle beams, which shows the convergence toward the expected RJ distribution $n_p^{\rm RJ}$ as $Q$ increases. 

%\textcolor{red}{Je suggere de supprimer la suite, c'est une erreur de notre part qu'on ne peut pas justifier, on n'aurait pas du montrer cette figure dans le PRL (on triche, c'est comme si on mettait un prior concentre sur le resultat voulu dans une approche bayesienne). je suis d'avis qu'il ne faut pas en parler, on ne peut faire que s'enfoncer nous memes.}

%\blue{We finally note that a  good agreement between the RJ distribution and a single experimental realization can be obtained in special cases, such as the Fig.~7 in the Supplementary Material of Ref.\cite{PRL23}. These correspond to a rare case where the mode reconstruction algorithm converges to a local minimum of ${\cal D}_Q$, which is much lower than what is expected from the theoretical expectation. This is consistent with the fact that there is no proof of convergence to the absolute minimum for the Gerchberg-Saxton algorithm.}

%\newpage


\begin{thebibliography}{99}
\bibitem{note}
In the absence of a waveguide, RJ thermalization is not well defined. 
Consider an experiment realized in a 2D bulk medium, where the RJ distribution takes the form $n_k^{\rm RJ} = T/(\omega_k - \mu)$, with $\omega_k \sim k^2$ the dispersion relation of the NLS equation. The power $N=\int n_k^{\rm RJ} d^2k$ and the kinetic energy $E=\int \omega_k n_k^{\rm RJ} d^2k$ exhibit an ultraviolet divergence at thermal equilibrium \cite{NP12}. 
To see this, consider the behaviour of the RJ spectrum for large frequencies $n_k^{\rm RJ} \sim 1/k^2$, then $N\sim \int dk/k$, and $E\sim \int dk$, which are divergent integrals for $k\to \infty$. The finite number of modes of the waveguide introduces an effective frequency cut-off that regularizes such an ultraviolet divergence \cite{PRA11b,PRL20}.
\end{thebibliography}

\end{document}